\documentclass{rmf-d}
\usepackage{nopageno,rmfbib,multicol,times,epsf,amsmath,amssymb,cite}
\usepackage[T1]{fontenc} %Especial para español (for spanish)
\usepackage[]{caption2}
\usepackage{graphicx}
\usepackage{blindtext}
\usepackage{color}
\usepackage{hyperref}
\def\rmfcornisa{Research OR Education of Physics \hfill\rmf\ {\bf ??} (*?*) ???--???
\hfill MES? A\~NO?}

\def\rmfcintilla{{\it Rev.\ Mex.\ Fis.\/} {\bf ??} (*?*) (????) ???--???}
\clearpage \rmfcaptionstyle 
\newcommand{\pp}{pp\xspace} 
\newcommand{\ppb}{p--Pb\xspace}

\newcommand{\pbpb}{Pb--Pb\xspace}

\newcommand{\ee}{\ensuremath{e^+e^-}\xspace}
\newcommand{\eebold}{\ensuremath{\mathbf{e^+e^-}}\xspace}
\newcommand{\qqbar}{\ensuremath{q\bar{q}}\xspace}

\newcommand{\pA}{pA\xspace}
\newcommand{\AAA}{AA\xspace}

\newcommand{\npart}{\ensuremath{N_{\text{part}}}\xspace}

\newcommand{\pT}{\ensuremath{p_{\text{T}}}\xspace}

\newcommand{\dndeta}{\ensuremath{\text{d}N_{\text{ch}}/\text{d}\eta}\xspace}

\begin{document}
\markboth{ RMF Editorial Team    }{ A \LaTeX template for the RMF, RMF-E, SRMF }

%%%%%
%
% Please provide the following information
%
%%%%%
\title{Perfect QCD -- a new Universal approach to soft QCD
\vspace{-6pt}}
\author{P. Christiansen}
\address{Division of Particle Physics, Lund University, Sweden}
\maketitle
%%%%%
%
% To be filled by the Editorial Team of RMF, RMF-E 
% and SRMF
%
%%%%%
%\recibido{15 April 2022}{16 April 2023
%\vspace{-12pt}}
\begin{abstract}
\vspace{1em} 
%%%%%
%
% Provide your abstract
%
%%%%%
The ideas presented in this proceeding aims to be a first step towards a
description of hadronic collisions where all soft processes are fundamentally
strongly coupled and the same Universal strongly coupled physics drives both
initial and final-state interactions. \\ As it is not currently possible to
derive such a picture from first principles, instead, an attempt to generalize
the perfect liquid observation to a ``perfect QCD'' guiding principle is
presented, focusing on implications for particle production in small
systems. The first steps towards a microscopic model is taken by arguing that
``perfect QCD'' suggests that the screening in the initial state is so large
that multi-parton interactions are of little or no importance. Instead, a
target and projectile remnant is coherently excited and particle production is
mainly driven by radiation in a qualitative similar manner as $\ee \rightarrow
\qqbar$. \\Finally, some of the possible implications of this ``excited
remnant model'' are presented. It is argued that the time ordering of soft and
hard physics can explain the absence of jet quenching in small systems and
that the coherence scale of the projectile and target provides insights into
what small systems will exhibit flow.
\vspace{1em}
\end{abstract}
\keys{ \bf{\textit{%Provide Keywords
}} \vspace{-8pt}}
\begin{multicols}{2}

\section{Introduction}
\label{sec:intro}

The goal of this proceeding for the Winter Workshop 2022 is to present a new
picture for hadronic collisions. To be precise, the focus in this paper is
only on non-diffractive inelastic collisions and only the soft
physics\footnote{Meaning that momentum transfers are small so that
  perturbative calculations are inaccurate. As, for example, the Pythia
  generator~\cite{Sjostrand:2014zea,Sjostrand:2006za} for proton-proton
  collisions treats all interactions perturbatively, this is not a unique
  definition but part of the motivation for exploring a completely different
  picture in this paper.}, which is expected to be responsible for bulk
particle production. When hadronic collisions are mentioned in the following
it always refers to this type of collision unless another type is explicitly
mentioned. \\ The motivation for doing this is the observation of several
phenomena in small systems\footnote{Small systems are
  taken to mean proton-proton, proton-nuclei and ultra-peripheral
  nuclei-nuclei collisions.} that has traditionally been associated with the
formation of a quark-gluon plasma (QGP) in large systems, see, e.g.,
Refs.~\cite{Loizides:2016tew,Nagle:2018nvi} for an overview. These new
phenomena can all be explained by the presence of large final-state
interactions in small system and many excellent ideas have been
presented for describing this with weakly coupled physics, see
e.g.,~\cite{Kurkela:2018xxd}, but what seems to the author to be a fundamental
flaw in these models is that a weakly coupled interaction leads to a
non-vanishing mean free path so that the QGP-like effects will build up as the
system grows and first dominate at a certain system
size~\cite{Kurkela:2018xxd}. This means that QGP-like effects do not in a
natural way extend down to the smallest systems, even if there is no
indication in data of an onset~\cite{Loizides:2016tew,Nagle:2018nvi}. At the
same time, a non-vanishing mean free path will introduce diffusion and
dissipation effects that will supposedly modify the initial-state
correlations, which the author is unaware of experimental evidence for, see
e.g. C.\ A.\ Pruneau's contribution to these proceedings~\cite{Pruneau:2022pkq}.

\begin{figure}[H]
  \begin{center}
  \includegraphics[width=\linewidth]{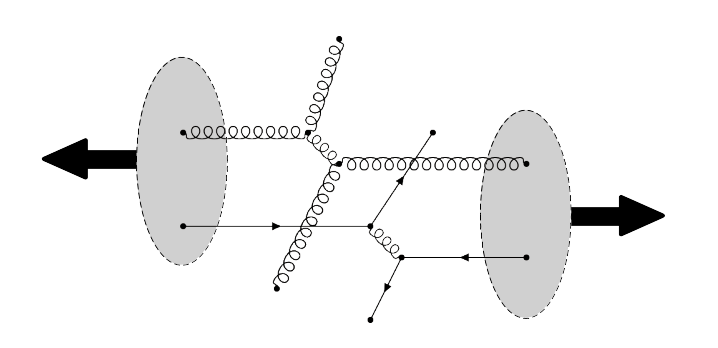}
  \includegraphics[width=\linewidth]{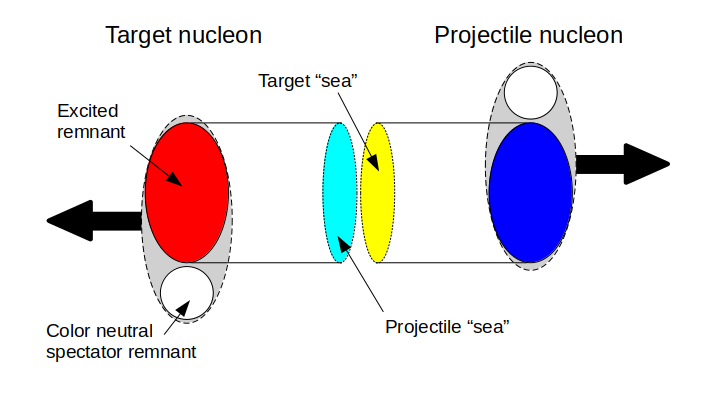}
   \caption{Illustration of how the initial soft scatterings are described in
     different models/pictures. Top: in Pythia, soft and hard interactions are
     modeled in the same way as leading-order perturbative processes and
     \emph{multiple} interactions occur in most \pp collisions. The strings
     forming between color charges are not shown. Bottom: in the ``perfect
     QCD'' picture a remnant of each nucleon is excited as a whole, as if
     there was only a \emph{single} interaction, and the picture is therefore
     denoted the ``excited remnant model''.}
  \label{fig:models}
\end{center}
\end{figure}

In this paper, the decision has been to take a fresh look at things from the
perspective offered by the new measurements and try to bring forth a picture
that is fundamentally strongly coupled with a vanishing mean free path so that
large final-state effects are present in all systems and do not introduce
diffusion or dissipation (are essentially reversible) thereby hopefully
preserving correlations such as those introduced by string breakings or
similar processes. In traditional pictures, ``soft'' can have two very different
meanings:
\begin{enumerate}
\item The extrapolation from high-momentum transfers to low momentum transfer,
  e.g., using leading-order perturbative cross sections even for situations
  where next-to-leading order correlations are large
\item Phenomenological physics such as the Lund string
  model~\cite{Andersson:1983ia}
\end{enumerate}
The approach in this paper is to claim that point 1 does not work, meaning
that next-to-leading order corrections distorts the leading-order picture, and
the proposal is instead that ``perfect QCD'' is a Universal version of point 2
and can provide guidance in that way. This means that any time where soft is
mentioned in the text one should in principle be able to apply the ``perfect
QCD'' principle. To help convince the reader that this leads to fundamentally
different physics from that found in existing models, one of the main findings
will be already discussed here and illustrated in Fig.~\ref{fig:models}. In pp
event generators, such as Pythia~\cite{Sjostrand:2014zea,Sjostrand:2006za},
one typically treats the initial stages of \pp collisions as two interacting
parton gases where the scattering of each parton-parton interaction is
motivated by perturbative (weakly coupled) QCD, Fig.~\ref{fig:models} top. In
the Color-Glass Condensate (CGC) model, not shown, one instead considers it as
a weakly coupled interaction between dense gluon fields~\cite{Gelis:2010nm}
that produce longitudinal Glasma tubes, Fig.~\ref{fig:models}. In both models
the collision can involve one or more interactions and \emph{the number of
  interactions is the main driving mechanism of the final-state
  multiplicity}. In the picture motivated in this paper, one considers a
strongly coupled scenario where the color field of each projectile parton is
neutralized by the target partons. It is argued that this results instead in
that the remnant of the projectile and the target is coherently excited,
corresponding essentially to a \emph{single} soft interaction. This gives rise
to two semi-independent color fields, Fig.~\ref{fig:models} bottom, which
would mean that most of the particle production is driven by final-state
radiation from the colored target and projectile remnants, similar to $\ee
\rightarrow \qqbar$.

Concretely, the idea of this paper is to extend the experimental observation
that the QGP behaves like a perfect liquid to a ``perfect QCD'' principle that
can guide our understanding of particle production in general. The goal is not
to come up with a full model, but to demonstrate that it is possible using the
proposed ``perfect QCD'' principle to obtain surprising insights into particle
production where the physics and the explanations for observed phenomena are
very different from those found in existing models, such as Pythia and the CGC.

\section{Perfect QCD}

One of the most remarkable discoveries of the heavy-ion program at RHIC and
LHC is that the Quark-Gluon Plasma (QGP) behaves as a perfect
liquid~\cite{Arsene:2004fa,Adcox:2004mh,Back:2004je,Adams:2005dq,Aamodt:2010pa,ATLAS:2011ah,Chatrchyan:2012ta}. The
shear-viscosity-to-entropy density ($\eta/s$) is as low as
possible~\cite{Kovtun:2004de}. This means that the build up of flow is almost
deterministic, which has enabled the precise measurement of fluctuations in
the initial distribution of matter, e.g.,
Refs.~\cite{Alver:2008zza,Alver:2010gr}. At the Winter Workshop it was further
shown how the same minimal $\eta/s$ is also obtained when analyzing balance
functions and momentum correlations, see C.\ A.\ Pruneau's contribution to these proceedings~\cite{Pruneau:2022pkq}.

The perfect nature of the liquid
seems to indicate that it is very fundamental and since it is observed in all
hadronic collisional systems (\pp, \ppb, and \pbpb collisions), see for
example Refs.~\cite{Khachatryan:2015waa,Aaboud:2017blb} for small systems, one
could hope that it provides a deep insight into QCD.\\

Based on the characteristics of the perfect liquid it is proposed that
``perfect QCD'' has to have the following two characteristics:
\begin{itemize}
\item Strongly interacting
\item Minimal entropy production
\end{itemize}
The minimal entropy production comes from the observation that the
hydrodynamic description of the QGP is as close to ideal (reversible) as it
can be and means that dissipation and diffusion can play no significant role
in the description of the system.

\section{The Perfect QCD Picture of Particle Production}
\label{sec:minimal}

It might seem impossible to derive a microscopic picture from a strongly
interacting soft QCD model because one looses the perturbative guidance but
the surprise is that the proposed picture is extremely simple. The ``perfect
QCD'' principle dictates that the entropy production during the initial
collisions should be as small as possible, yet strongly interacting, and this
suggests that all that happens is the exchange of a single soft gluon so that
only color and essentially no momentum is exchanged. As the interacting
hadrons are of course made up of partons, this would require that the
screening in the initial state is so strong for the initial interactions that
the soft parton-parton (and/or CGC equivalent) interactions are suppressed to
a degree where they can be neglected. One will of course have parton-parton
interactions for very large momentum transfers but they are not of interest
here where the focus is on bulk production.

Let us first treat the rest of the collision, ignoring possible radiation,
using the Lund string model~\cite{Andersson:1983ia}, which, as it is derived
from the confining long-range part of the QCD potential, is a strongly coupled
model. In the Lund string model, strings will form between colors and
anti-colors that eventually breaks, producing hadrons uniformly in
rapidity. In this case, two strings will form as the gluon carries both a
color and anti-color. Let us assume that all the energy of each proton is
carried by the color and the anti-color systems. If both have half the energy,
the total string length will be ${\approx}4(y_{\rm beam} - \log{2})$ while if
one color (or anti-color) has all the energy one can supposedly form a string
of length $2y_{\rm beam}$ (this must be the minimal length for the color field
to stretch between the target and projectile). As the average number of
particles produced a by a string is proportional to the string
length~\cite{Andersson:1983ia}, the ``perfect QCD'' principle tells us that
nature will take the 2nd solution. This means that instead of having two
remnants with a similar amount of energy, one will have a ``valence''-like
remnant with almost all the energy and a ``sea''-like remnant with almost no
energy. This is reminiscent of the BGK picture~\cite{Brodsky:1977de}, and so
it is naturally to propose that the ``valence'' remnant in one proton is
color-coupled to the ``sea'' remnant in the other proton, and vice versa, so
that one in some sense has two semi-independent systems carrying approximately
half the total initial energy each. \\ Let us finally try to give a partonic
picture of how the ``perfect QCD'' picture can be understood. As the two
nucleon penetrate at high energy the partons inside them are interacting
strongly but the claim is that they interact in a way that screens the
partonic interactions. However, this screening can only happen in a certain
regime. If $x$ denotes the usual four momentum fraction then one can maximally
``organize'' the nucleon into $n \approx 1/x$ constituents. Screening will be
impossible when the four momentum transfer, $Q^2$, is very large because one
can resolve individual partons (the hard scattering limit), or when $x$ is
large so that the number of constituents is small. The latter argument is why
nucleon remnants will be excited as a whole.

In the current picture, \dndeta at $\eta = 0$ would be independent of
$\sqrt{s}$ as all the energy will go to extend the strings in rapidity. What
has been ignored is radiation: the color charge carrying most of the energy
is, as QCD is strongly interacting, very likely to emit soft or collinear
radiation. How to calculate this radiation is not trivial, but one can at
least note that one qualitatively get a system very similar to what one has
for $\ee \rightarrow \qqbar$ (denoted \ee in the following). Comparing
particle production in \ee collisions to that of \pp collisions, one finds
that the former produces \emph{more} particles on the
average~\cite{Back:2004je}. The common understanding is that it is possible
for part of the proton to escape as a color neutral object, taking away around
50\% of the energy~\cite{Back:2006yw}. Based on the observed particle production in \ee, it is concluded
that there is no fundamental reason one should not be able to create the
observed particle production via radiation also in \pp, \pA, and \AAA
collisions. \\

To recap, the general microscopic ``perfect QCD'' picture of \pp, \pA, and
\AAA collisions will be that the soft initial interactions will excite a
remnant of each nucleon in a ``projectile'' coherently and that the main
particle production at high energy collisions is driven by final-state
radiation. For this reason the picture will be denoted the ``excited remnant
model''. This might sound like the Dual Parton Model but it is important to
note that the Dual Parton Model contains MPIs~\cite{Capella:1992yb}. \\

In the limit that particle production is dominated by radiation, the
color-connections to the ``sea'' systems in the ``target'' can be ignored and
one can therefore factorize the soft particle production into \npart
semi-independent terms. Semi-independent, because there must be some
dependence on the nucleon-nucleon impact parameter to explain the slightly
increased particle production per participant in \AAA collisions.\\

\subsection{An Illustration of Particle Production in \pp Collisions}

\begin{figure}[H]
  \begin{center}
  \includegraphics[width=\linewidth]{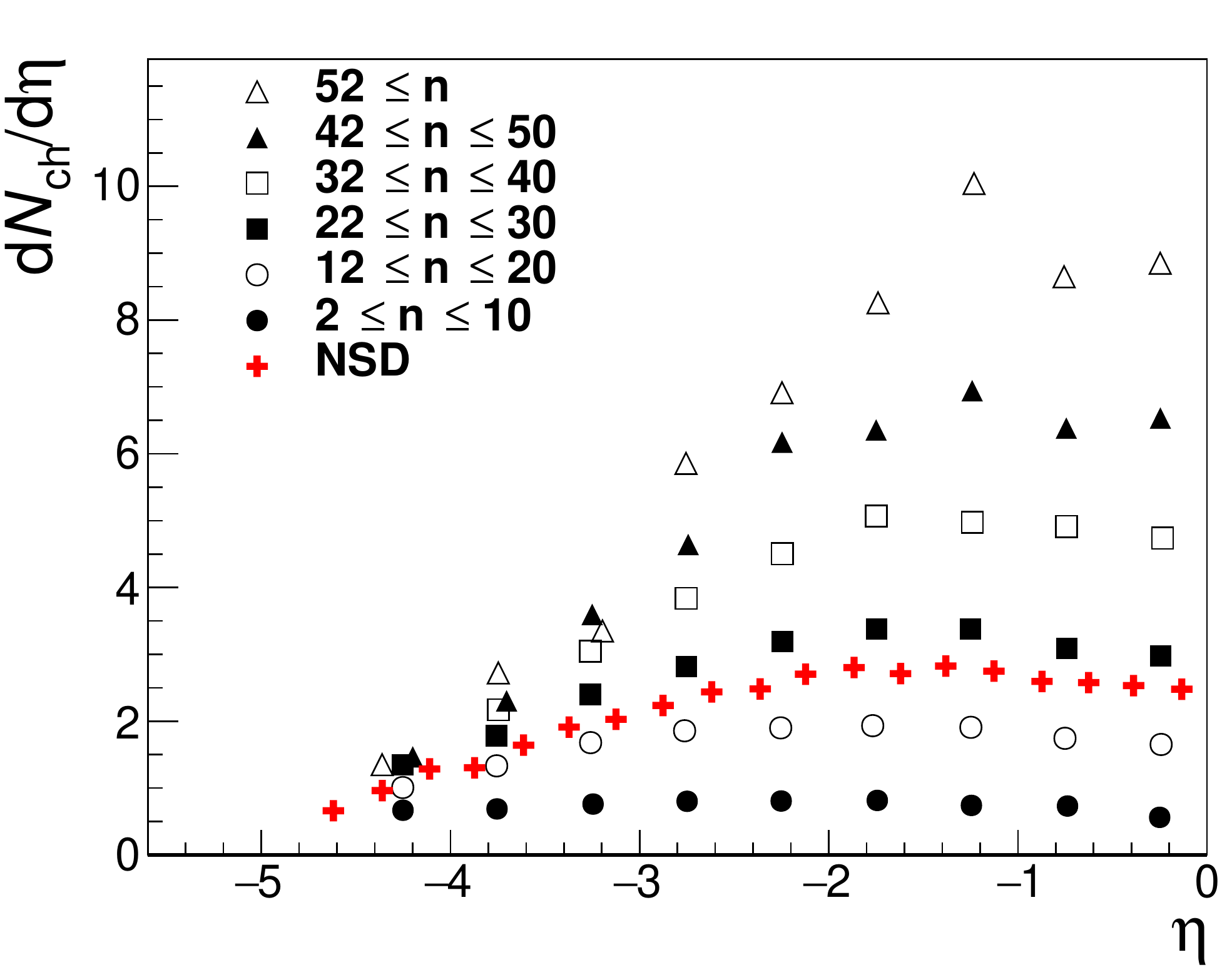}
   \caption{\dndeta measured in $\sqrt{s} = 200$\,GeV/$c$ \pp collisions by
     UA5 for NSD events and for events with different final-state charged
     particle multiplicities, $n$. The data have been read off from the
     published figures~\cite{UA5:1986yef}. As the figure is just meant to
     illustrate a trend, the statistical uncertainties have not been included
     for clarity.}
  \label{fig:ua5}
\end{center}
\end{figure}

The main goal here is to discuss small systems. In these systems,
e.g., pp collisions, the full ``perfect QCD'' picture of a collision is:
\begin{enumerate}
\item the initial interactions produce up to three semi-independent systems:
  \begin{itemize}
  \item coherently excited target and projectile remnants
  \item possible color-neutral target and projectile remnants that act as spectators (escape with energy along beam direction) 
  \item possible hard parton-parton scatterings
  \end{itemize}
\item the excited remnants radiate gluons
\item the color fields decay into partons
\item final-state partonic interactions: flow, strangeness enhancement
\item hadronization
\item possible final-state hadronic rescattering
\end{enumerate}

One could in principle try to implement a generator along these lines but the
goal here is to illustrate the picture using UA5
data~\cite{UA5:1986yef}. Fig.~\ref{fig:ua5} shows the \dndeta measured by UA5
for NSD events as well as for multiplicity selected events. In
low-multiplicity events, \dndeta is flat as one would expect for a single long
string. As the multiplicity grows, one observes a narrowing of \dndeta, which
in the perfect QCD picture should be caused by the radiation adding shorter
and shorter (less energetic) strings. In this way the ``excited remnant model''
is at least qualitatively consistent with the observed trends by UA5.

\section{Insights and Predictions for Small Systems}

In this section, the hope is to demonstrate for the reader that the
perfect-QCD picture of particle production can provide many new insights and
predictions.

\subsection{A Simple Explanation for the Absence of Jet Quenching in Small Systems}

One can immediately notice that, if the time scales involved with the hard
interactions are shorter than the formation time for step 2 (``the excited
remnants radiate gluons'') as one would imagine from the scales of the
momentum transfers involved, then one can understand why there is no jet
quenching in small systems even if there is a relation between flow and jet
quenching in a large system. The medium simply has not been produced yet when
the jet propagates. This seems very attractive to the author as this is in
line with experimental findings, see, for example, Ref.~\cite{ALICE:2017svf},
and it is hard to explain in most existing models.

\subsection{Flow in \pp Collisions $\gg$ Flow in \eebold Collisions}

It should be clear from the way the ``excited remnant model'' work that it
``postdicts'' that the particle production in \pp collisions and \ee
collisions should be very similar because in this model, and unlike traditional
MPI-based models, the growth with $\sqrt{s}$ is in both cases driven by
radiation. Indeed this surprising similarity have been noted and discussed
much in the past by experimental
collaborations~\cite{Zichichi:2011zz,Back:2004je}, even it was never
theoretically understood.

It can therefore be surprising that while one observes strong flow in pp
collisions, one does not observe it for \ee
collisions~\cite{Badea:2019vey}. However, there could be a simple explanation
for that. As the ``excited remnant model'' postulates that for each nucleon a
single ``valence'' remnant is excited as a whole, then it is clear that the
radiation in step 2 will have to have very low transverse momentum, $\pT <
1/R$, where $R$ is the size of the excited remnant. As the \pT is so low, the
color fields will have to stack and so one will naturally get a quite dense
system of parallel color fields with a large energy density. In the ``perfect
QCD'' picture these color fields will be strongly interacting and so they will
immediately start to build up collective flow. This makes a big difference
when comparing to \ee, where all the energy is located with a single parton
and so the radiated gluons can and will typically have very large \pT. This
means that most energy will be radiated away from the initial color field and
so there is little time where system is dense and can build up collective
flow.

\subsection{How to Control Flow in Ultra-Small Systems}

In the previous subsection it was argued that for small systems, the size of
the excited remnant determines the flow that can be built up in the final
system. This then is naturally in line with the observation of flow in
Ultra-Peripheral Collisions (UPCs), where the photon field of one nuclei
interacts with the other nuclei, because in this case the photon field has a
long wavelength since it is emitted coherently by the protons in the
nuclei. Recall that photons can interact as a ``hadronic'' system by
fluctuating into a \qqbar pair, which will have a size that reflects the
photon four momentum ($Q^2$). ATLAS has observed flow in UPC \pbpb
events~\cite{Aad:2021yhy} and CMS has reported non-zero $v_2\{2\}$ in \ppb
events~\cite{CMS:2020rfg}, which is in line with the ideas presented here.

By going to electron-proton or electron-ion collisions one can in principle
measure the wavelength of the photon from the change in electron four
momentum. One can in this way select different sizes of the excited remnants
and if the picture is true, control the \pT radiation and switch on (low
$Q^2$) and off (high $Q^2$) flow. ZEUS and H1 has reanalyzed old data both for
low and high $Q^2$ but neither ZEUS~\cite{ZEUS:2019jya,ZEUS:2021qzg} nor
H1~\cite{H1} observes any signatures of collective flow. This clearly goes
against the ideas presented here. However, it seems that if there is flow in
UPCs at LHC then there would also likely be flow in low $Q^2$ $e$p collisions
at HERA and vice verse. On the other hand, one knows that flow in small
systems is very hard to detect. Looking from the outside, it would be good if
one could resolve the situation so that one is as certain as possible that
similar procedures have been used before one concludes too strongly on the
current results.

\section{Conclusions}
    
An attempt to generalize the perfect-liquid nature from flow to particle
production has been presented. The ``perfect QCD'' principle has been proposed
to be a Universal principle for soft QCD that applies both in the initial and
final state of hadronic collisions. Using the idea of minimal entropy
production, a microscopic picture, the ``excited remnant model'', has been
presented. In the microscopic picture, the screening as the two hadronic
systems penetrate is so large that subcollisions between constituents does not
occur, in contrast to most existing pictures, e.g., MPI and CGC based ones.

No attempt has been done to prove the ``perfect QCD'' principle in this paper
but several surprising insights have been provided, such as simple arguments
for why jet quenching is absent in small systems and which collisional systems
will exhibit flow. The hope is that the principle can be used to provide novel
insights into a wide range of topics, for example, jet quenching in large
systems and the relation between diffractive and non-diffractive physics.

\section{Acknowledgements}

The author would like to thank Adrian Nassirpour for many valuable comments
on earlier versions of similar manuscripts.

\end{multicols}

\section{References}

\medline
\begin{multicols}{2}
%%%%%%%%%%%%%%%
%
%Using BibTeX
\nocite{*}
\bibliographystyle{rmf-style}
\bibliography{bibliography}
\end{multicols}
\end{document}